\documentclass[preprint,floatfix,aps,prd,showpacs,footinbib,amsmath,amssymb,amsfonts,superscriptaddress]{revtex4-1}

\usepackage{amsmath}    
\usepackage{graphicx}   
\usepackage{verbatim}   
\usepackage{color}      

\usepackage{subfigure}  
\usepackage{hyperref}   
\hypersetup{
	colorlinks=true,       
	linkcolor=blue,          
	citecolor=blue,        
}

\usepackage{dsfont}
\usepackage[normalem]{ulem}
\usepackage{mathtools}
\usepackage{mathrsfs}
\usepackage{calrsfs}
\usepackage{comment}
\usepackage[utf8]{inputenc}
\usepackage{varioref}
\usepackage[active]{srcltx}

\expandafter\ifx\csname package@font\endcsname\relax\else
\expandafter\expandafter
\expandafter\usepackage
\expandafter\expandafter
\expandafter{\csname package@font\endcsname}%
\fi

\usepackage[section]{placeins}

\usepackage{fancybox}
\labelformat{figure}{Fig.~#1}

\begin{document}

\title{Detectability of nGauss primordial magnetic fields in the B-mode polarization of CMB}

\author{Archana Sangwan}
\email{arch06san@gmail.com}

\author{S. Shankaranarayanan}
\email{shanki@phy.iitb.ac.in}
\affiliation{Department of Physics, Indian Institute of Technology Bombay, Mumbai 400076, India}

\begin{abstract}
The origin of large-scale magnetic fields in the Universe is still unknown. Observations suggest the presence of nGauss primordial magnetic fields at the last scattering surface.  The presence of such fields will affect the evolution of cosmological perturbations and potentially leave an imprint on the CMB anisotropies. Here, we show that the B-mode power spectrum carries a clear signature of the stochastic primordial magnetic fields up to a few nGauss. Specifically, the presence of nGauss primordial magnetic fields changes the BB power spectrum at all scales. 
At large scales, the tensor modes contribute to the B-mode 
spectrum, while non-vanishing vector modes contribute at small scales. We show that the B-mode of CMB carry a distinct signature of the primordial magnetic fields. 
To validate our prediction, we use BICEP2 and POLARBEAR data and find that B-mode observations from the experiments are consistent with non-zero primordial fields. We also use the BICEP2 observations to constrain the primordial magnetic field. We provide a detailed analysis of the B-mode polarization of the CMB with primordial magnetic fields and investigate the implications for the upcoming CMB missions.
\end{abstract}

\keywords{Magnetic fields, CMB, B-modes}
\maketitle

\section{Introduction}
\label{sec::intro}

Observations indicate that the Universe is filled with magnetic fields at
all length scales~\cite{2015-Rainer-AAR,2016-Subramanian-RPP}.
Galactic fields of micro-Gauss strength and intergalactic magnetic fields of strength $10^{-7}-10^{-6}~\rm{G}$ are observed~\cite{Bernet:2008qp}.
While the magnetic field measurements from Faraday rotation~\cite{2009-Taylor.etal-ApJ}, and synchrotron radation~\cite{2004-Govoni.Feretti-IJMPD} provide upper bounds of the magnetic fields, the FERMI measurement of
gamma-rays emitted by blazars seem to provide lower bound of the order of $10^{-15}~\rm{G}$ in voids~\cite{2010-Neronov.Vovk-Science,*2010-Tavecchio.etal-MNRAS,*dermer2011time}. 

Like the growth of large scale structures, magnetic fields can
be amplified either by the dynamo mechanism which requires a tiny seed
fields of the order $10^{-20} - 10^{-30}~\rm{G}$ or adiabatic compression of a primordial magnetic field, where a larger seed field with an amplitude of about $10^{-9} - 10^{-10}~\rm{G}$ to generate the fields we observe today~\cite{1979cmft.bookParker,*1983mfa.bookZeldovich}.
The observations of magnetic fields at high redshift have raised 
questions about the effectiveness of the dynamo mechanism~\cite{ko1989,*882457b272fb458ea1aa77f568c656a9,*PhysRevLett.83.2957}.
First, the mean-field dynamo, responsible for large-scale magnetic fields, would not have enough time to amplify the seed fields to the observed values in high redshift galaxies \cite{wolfe1992}. Second, the observation of the absence of the photons in ${\rm GeV}$ range from the ${\rm TeV}$ Blazars suggests the presence
of non-zero magnetic fields in voids and sets a lower limit to the magnetic field~\cite{2010-Tavecchio.etal-MNRAS,2010-Neronov.Vovk-Science}. 
However, the turbulent plasma dynamo can not operate in voids, hence, 
the magnetic field in the voids must reflect the primordial seed field (before the amplification).

This strongly suggests that a non-astrophysical process is responsible for the magnetic fields in the voids. One such alternative is to assume that these large scale magnetic fields resulted from some primordial magnetic field generated in the early Universe~\cite{2016-Subramanian-RPP}. A primordial magnetic field in the early Universe can explain the magnetic fields observed on large scales as these fields can be amplified to the present observed value in galaxies and galaxy clusters, but remain mostly unaffected in the voids by different astrophysical processes over the evolution time of universe~\cite{Hutschenreuter:2018vkr}.

The conductivity of the intergalactic medium during most of the 
history of the universe was high, hence the magnetic fields are frozen, and the magnetic flux is conserved. Due to the cosmological expansion, the magnetic field strength decreases, i.e. 
\begin{equation}
B(t_2) = \left (\frac{a(t_1)}{a(t_2)}\right )^2
B(t_1 ) = (1 + z)^2 B(t_1) 
\end{equation}
where $t_2 > t_1$. Given that the magnetic field strength in the voids (at low redshifts) is of the order $10^{-16}~{\rm G}$, the above relation leads to the fact that magnetic field at the time of the Last Scattering surface (LSS) is of the order $10^{-9}~ {\rm G}$. This is consistent with the constraints obtained using different methods, such as the non-Gaussianity of the temperature fluctuations of CMB provide an upper limit of $35~{\rm nG}$ on primordial magnetic field~\cite{2009-Seshadri.Subramanian-PRL,2014-Bonvin.etal-PRL}.

Faraday rotation provide a limit of $B\sim10^{-9} ~ {\rm G}$ within 3-sigma level~\cite{Kolatt:1997xu},
CMB anomalies give an upper limit of $15~ {\rm nG}$ within 3$\sigma$ confidence level~\cite{Bernui:2008ve}, and effects on large-scale structure (LSS) constrain magnetic field to $10^{-8}-10^{-9}$G \cite{10.1046/j.1365-8711.2003.06610.x,*Sethi:2004pe},
and the end of reionization provide the upper limit on primordial magnetic field to be $2-3~ {\rm nG}$~\cite{2011MNRAS.418L.143S}.
CMB observations provide strong direct limits on the field strengths of several ${\rm nG}$~\cite{PhysRevD.81.103519,*PhysRevD.78.123001}.

The generation of primordial magnetic fields (PMF) in the early Universe can change the evolution history of the Universe as these source scalar, vector, and tensor modes of the cosmological perturbations. In the context of observations, these three modes will modify the evolution of temperature and polarization perturbations. They leave an imprint on the temperature and polarization anisotropies power spectrum in CMB~\cite{1996-Adams.etal-PLB,2004-Lewis-PRD,PhysRevD.72.103003,10.1111/j.1365-2966.2009.14727.x,2010-Shaw.Lewis-PRD}.

In this work, we address the question of detectability of nGauss PMF using the $B$-mode power-spectrum of the CMB.  We explicitly show that 
$B$-mode power-spectrum of the CMB carries a clear signature of the PMF up to a few ${\rm nGauss}$. The only assumption we make about the PMF is that it is stochastic. We do not make any assumptions about the origin of the PMF and could have been produced during inflation or later (causal field). (See, for instance, Refs.~\cite{1988-Turner.Widrow-PRD,*PhysRevLett.75.3796,
*2015-Basak.Shankaranarayanan-JCAP,*2016-Fabre.Shankaranarayanan-APP,*2018-Nandi.Shankaranarayanan-JCAP}.) While the effect of PMF on the \emph{TT, TE}, and \emph{EE} power-spectrum is negligible~\cite{2004-Lewis-PRD,2010-Shaw.Lewis-PRD}, there is a significant change in the $B$-modes of the CMB power spectrum.  Specifically, there is an increase in the \emph{BB} power spectrum at all $\ell$ due to nGauss PMF. 

In the past, attempts have been made to remove lensing effect on the anisotropies spectrum~\cite{Green:2016cjr}. 
It is well-known that the lensing mixes $E$ and $B$ polarization modes. Even for pure scalar fluctuations, B-modes are generated at small scales~\cite{Millea:2017fyd}. This is because a large number of astrophysical processes contribute to the $E$-mode, and hence, lensing induces an increase in the $BB$ power-spectrum at large $\ell$. However, PMF introduces vector perturbations leading to a rise in the \emph{BB} power-spectrum at large $\ell$. We also show that lensing does not contribute significantly to the power-spectrum for PMF greater than $3~{\rm nG}$. To our knowledge, this is the first time the B-mode of CMB is used to measure the PMF in the recombination epoch. To validate our prediction, we use BICEP2~\cite{Ade:2014xna} and POLARBEAR~\cite{2017-POLARBEAR-ApJ} data to see if the power spectra with a non-zero nGauss PMFs are consistent with the observations. We also obtain the allowed range of PMF strength with which the BICEP2 observations are consistent.

In Section \ref{sec::PMF}, we provide the fundamental equations and the quantities that are used to describe PMF. In Section \ref{sec::power_spectrum}, we discuss its effects on the power spectrum of temperature and polarization anisotropies of the CMB. 
We show how the B-modes can be used to detect the primordial magnetic fields. In Section \ref{sec::results_and_conclusions}, we discuss the results and implications for future CMB missions.

\section{Primordial magnetic field and CMB power spectra}
\label{sec::PMF}

As mentioned earlier, we assume that the primordial magnetic field 
($\mathbf{B}(\eta, \mathbf{x})$) is stochastic. Since the conductivity of the intergalactic medium is effectively infinite, the scaling of PMF is given by 
$\mathbf{B}(\eta, \mathbf{x})=\mathbf{B}_{0}(\mathbf{x}) / a^{2}(\eta)$, 
where $\eta$ is the conformal time and $a(\eta)$ is the scale factor.
The PMF power spectrum which is defined as the Fourier transform of the two-point correlation is~\cite{Durrer_2003}:
\begin{equation}
\left\langle B_{l}(\mathbf{k}) B_{m}^{*}\left(\mathbf{k}^{\prime}\right)\right\rangle=(2 \pi)^{3} \delta^{3}\left(\mathbf{k}-\mathbf{k}^{\prime}\right)\left(P_{l m}(k) P_{B}(k)+i \epsilon_{\operatorname{lm} n} \hat{k}^{n} P_{H}(k)\right)
\end{equation}
where $P_{l m}(k)=\delta_{l m}-\hat{k}_{l} \hat{k}_{m}$ is a projector onto the transverse plane, i.e., $P_{l m} \hat{k}^{m}=0$ with $\hat{k}=\hat{k}$ and $\epsilon_{l m n}$ is the $3 \mathrm{D}$ Levi-Civita tensor. $P_{B}(k)$ and $P_{H}(k)$ are the symmetric and antisymmetric parts of the power spectrum and represent the magnetic field energy density and absolute value of the kinetic helicity, respectively:
\begin{eqnarray}
\begin{aligned}\left\langle B_{i}(\mathbf{k}) B_{i}^{*}\left(\mathbf{k}^{\prime}\right)\right\rangle &= 2(2 \pi)^{3} \delta^{3}\left(\mathbf{k}-\mathbf{k}^{\prime}\right) P_{B}(k) \\-i\left\langle\epsilon_{i j} k^{l} B_{i}(\mathbf{k}) B_{j}^{*}\left(\mathbf{k}^{\prime}\right)\right\rangle &= 2(2 \pi)^{3} \delta^{3}\left(\mathbf{k}-\mathbf{k}^{\prime}\right) P_{H}(k) \, .
\end{aligned}
\end{eqnarray}
One usually assumes that the power spectrum scales as a simple power law, i. e.,
\begin{equation}
P_{B}(k)=A_{B} k^{n_{B}}, \quad P_{H}(k)=A_{H} k^{n_{H}}.
\end{equation}
where $A_{B} (A_{H})$ is the amplitude and $n_{B} (n_{H})$ is the spectral 
index. PMF can affect the evolution of the temperature and the polarization of the CMB~\cite{Jedamzik:1996wp,Subramanian:1997gi,2004-Banerjee.Jedamzik-PRD,2010-Shaw.Lewis-PRD,10.1111/j.1365-2966.2009.14727.x}.  The temperature and $E-$mode polarization power spectrum are most impacted by PMF at larger multipoles (small scales). In small scales, the scalar power-spectrum is suppressed due to Silk damping, however, the contribution from the PMF through the vector contribution dominates the scalar modes. $B$ modes are not sourced by 
scalar~\cite{2001-Seshadri.Subramanian-PRL,*2003-Subramanian.etal-MNRAS}. In the standard cosmology, without the primordial field, \emph{only} the tensor perturbations source the $B$-modes and scalar perturbations have no effect. Hence, the detection of $B$-mode is referred to as smoking gun for inflation~\cite{Ade:2014xna}. The presence of PMF implies the presence of 
non-vanishing vorticity and hence, non-decaying vector perturbations~\cite{2015-Basak.Shankaranarayanan-JCAP}. Thus, PMF affects the temperature, $E$ and $B$ modes of the CMB power-spectrum. 

To investigate the impact of a PMF on the CMB power spectra, we consider a non-zero PMF amplitude such that $A_B \neq 0$ and $A_H = 0$. More precisely, we choose three different PMF strength --- $2.0~{\rm nG}$, $2\sqrt{2}~{\rm nG}$ and $4.0~{\rm nG}$ --- at the comoving length of $1$ Mpc.

To study the effects of PMF on the CMB, we use the publicly available CAMB code~\cite{Lewis:1999bs,CAMB}. We use the flat $\Lambda$CDM model with parameters compatible with the Planck 2018 data as the fiducial cosmological model to obtain the CMB power  spectra~\cite{Akrami:2018vks}. We consider only massless neutrinos in this analysis. As mentioned earlier, the intensity of polarized radiation expressed in terms of $E$ and $B$ modes are independent of the orientation of the coordinate system. Thus, the CMB power spectrum has four observables: the temperature, E-mode, B-mode, and the temperature cross-polarization power spectra. From these, one can generate three power spectra TT, EE, and BB, and also cross-spectra. 

\begin{figure}[!htb]
\hspace*{-2cm}
    \centering    
\includegraphics[scale=0.38]{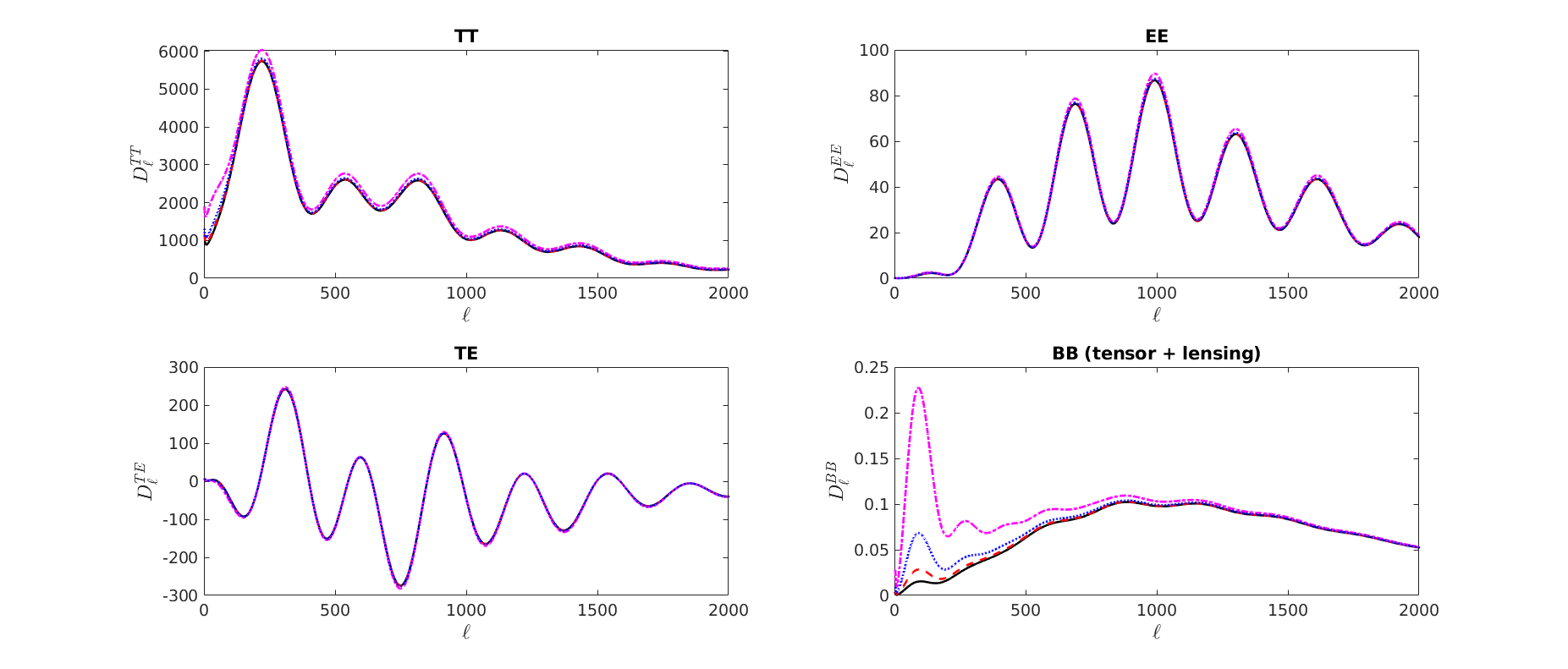}    
    \caption{The plot of the angular power spectra as a function of $\ell$ for vanishing and non-zero PMFs. In the first row, the plot on the left (right) is for TT (EE).  In the second row, the plot on the right (left) is for BB (TE). In these plots, we have considered \emph{only} the tensor mode and lensing contributions, and ignored the vector contribution. The black curve is for the zero magnetic fields, while the magenta, blue and red curves correspond to a PMF strength of $2.0$ nG,  $2\sqrt{2}$ nG and  $4$ nG, respectively at the comoving length of $1$ Mpc.} 
    \label{fig::tt_ee_te_bb}
\end{figure}

\ref{fig::tt_ee_te_bb} contains the plot of four angular spectra  --- TT (Top left), EE (Top right), TE (bottom left) and BB (bottom right)  ---  as a function of $\ell$  for the fiducial model. 
In this plot, we have suppressed the contribution of vector modes sourced by primordial magnetic fields and have considered the tensor modes and lensing contribution to the power spectra.
Different curves correspond to different values of PMF strength. The black curve corresponds to zero PMF, while the red, blue, and magenta curves are for non-zero nGauss PMF with increasing strength. 

We like to discuss the following points regarding \ref{fig::tt_ee_te_bb}: 
First, we find there is a little change in the TT, EE, and TE power-spectrum as the amplitude of the magnetic field is increased from zero to $4$ nG. Unlike the photons, magnetic fields are not affected by the diffusion; hence, for large multipoles $\ell$, the TT mode power spectrum is influenced by the PMF. However, the effect is not significant enough for nGauss range PMF.  Like the TT power spectrum, the effect is not significant enough for the EE and TE spectra. 
Second, in Ref. \cite{1996-Adams.etal-PLB}, the authors used $0.1~\mu{\rm G}$ PMF --- two orders of magnitude larger than used in this work --- and found changes in the $TT$ power-spectrum. However, as discussed earlier, the PMF strength of $0.1~\mu{\rm G}$ is inconsistent with other observations.
Third, in the case of the BB power spectrum, we see that as the magnetic field increases from zero value, the power in the spectrum increases. This is because PMFs sources the tensor modes which contribute at larger scales~\cite{2000-Durrer.etal-PRD}. The larger the strength of PMF is, the more is the contribution to the power spectra at smaller $\ell$. As $\ell$ increases, the PMF tensor contribution dies out, and it isn't easy to distinguish between the zero and non-zero PMF scenario.

\begin{figure}[!hbt]
\hspace*{-2cm}
    \centering   
\includegraphics[scale=0.37]{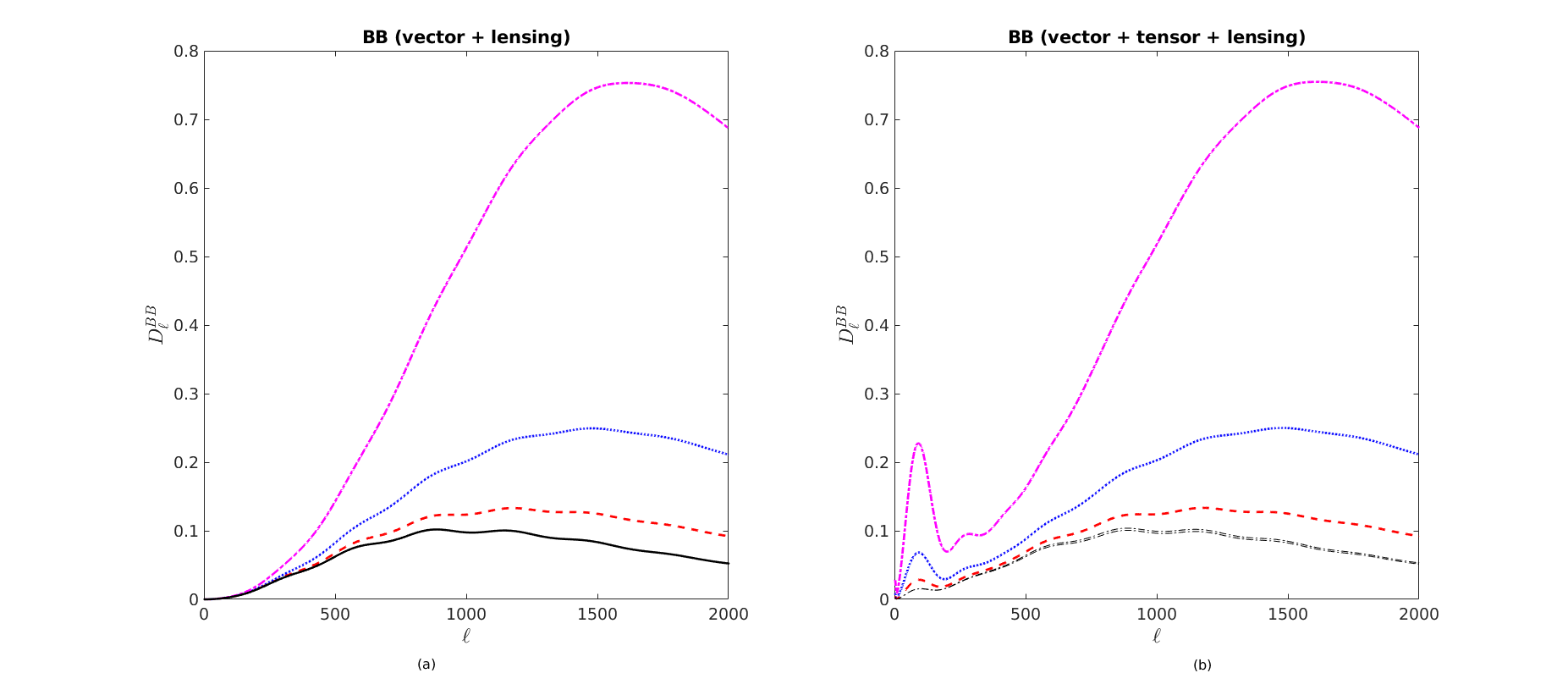}    
    \caption{The plot of the BB power spectrum as a function of $\ell$ for vanishing and non-zero PMFs. In the left plot, we have considered \emph{only} the vector modes and lensing contributions and ignored the tensor contribution. The plot on the right is the total BB spectra as a function of $\ell$. The colour scheme for curves is same as in \ref{fig::tt_ee_te_bb}. The black dotted lines in the plot (b) are obtained by varying cosmological parameters in the range consistent with Planck 2018 results for vanishing PMF. }
    \label{fig::bb_notensor_all}
\end{figure}

\ref{fig::bb_notensor_all} (a) contains the contribution of vector modes, along with lensing, on BB power spectra as a function of $\ell$ for different PMFs. The color scheme of the plot is the same as mentioned earlier. As can be seen from \ref{fig::bb_notensor_all} (a), the vector modes contribute at smaller scales, and the different curves corresponding to different PMF amplitude moves away from one other as $\ell$ increases. The maximum contribution to the power spectrum comes from the largest magnitude considered in work, followed by others in the decreasing order. Comparing this figure with that in \ref{fig::tt_ee_te_bb}, we see that the contribution of the vector modes substantially changes the BB power-spectrum at larger $\ell$. Thus, the lensed BB spectra sourced by the vector modes due to PMF can be used to detect nGauss PMF at the last scattering surface. 

\ref{fig::bb_notensor_all} (b) contains total B-mode power spectrum. The two dashed black curves represent the region within which the B-mode power spectrum lies, as long as there is no primordial magnetic field in the universe. This allowed range is obtained by varying the cosmological parameters according to the constraints reported in Planck 2018 paper \cite{Akrami:2018vks}. The colored lines show the BB spectrum for non-zero PMF with the same colour scheme as mentioned in \ref{fig::tt_ee_te_bb}. The above analysis clearly shows that the $B-$ mode of the CMB power-spectrum carries a distinct signature of the primordial magnetic field. However, this does not affect TT, TE, and EE power spectra. In the rest of the article, we will use the current data to constrain the nGauss primordial magnetic field. 

\section{Constraints from BICEP2 and POLARBEAR data}
\label{sec::power_spectrum}

BICEP2 (Background Imaging of Cosmic Extragalactic Polarization) and POLARBEAR experiments are designed to search the evidence of B-mode polarization in CMB signal~\cite{2014-BICEP2-ApJ,2017-POLARBEAR-ApJ}. 
While BICEP2 is sensitive to B-mode polarization on angular scales of $1-5^\circ$ ($\ell \sim 40-200$), POLARBEAR observed $25~{\rm degree}^2$  of effective area of sky with a resolution of $3.5'$ ($\ell \sim 500 - 2100$) at $150~GHz$. In the small multipoles, the B-mode signal induced by gravitational waves from inflation peaks, thereby the evidence for the detection of primordial gravitational waves from inflation. In the large multipoles, a $B$-mode signal is expected to be generated by gravitation lensing of $E$-mode polarization. Various detectors used in BICEP2 gave it unprecedented sensitivity to B-modes at these angular scales. Throughout the 3-year run of BICEP2, its detector biases and multiplexing rate had been optimized. These modifications improved the instantaneous sensitivity and mapping speed significantly, which gives an improvement in the map depth. Since it is a ground-based experiment, the residual atmospheric effects have put a limitation on its performance~\cite{Errard:2015twg}. Also, the point-like sources that have a spectrum similar to CMB contribute to the power-spectrum significantly~\cite{Bouchet:1999gq,*Giommi:2003xf}. So a careful cleaning for contamination of CMB maps is necessary~\cite{2014-PLANCK-AA}.  Extensive tests were performed to quantify the systematic and statistical errors in POLARBEAR B-mode signal~\cite{2017-POLARBEAR-ApJ}.

\begin{figure}[!hbt]
\hspace*{-2cm}
    \centering 
\includegraphics[scale=0.4]{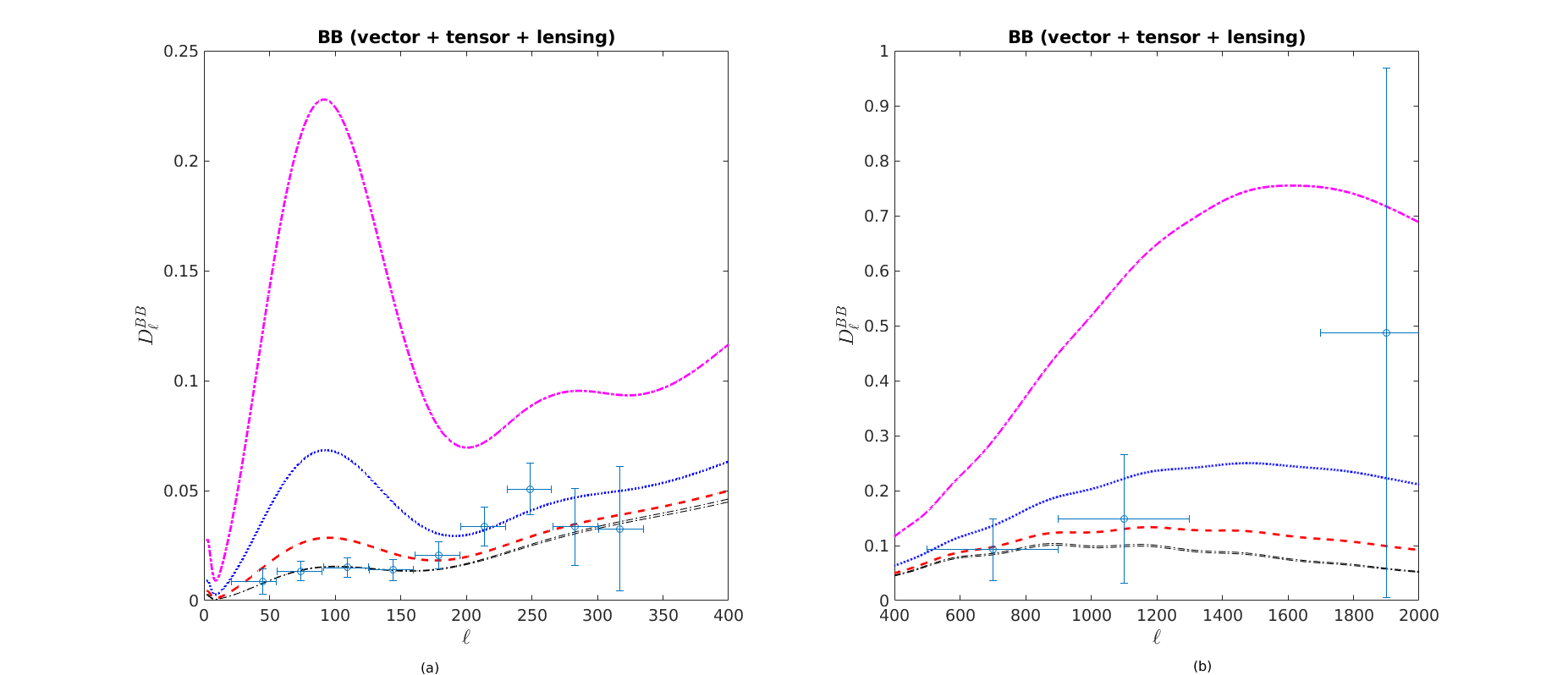}    
    \caption{The plots show the total BB angular spectra along with the observational data. The plot on the left shows the B-mode spectrum with the data points from BICEP2 observations with the error-bars, up to $\ell = 400$. The plot on the left shows the B-mode spectrum at higher $\ell$ along with POLARBEAR observations of BB-power spectra, with error-bars. The two black dotted curves show that the B-mode spectrum and its allowed range consistent with Planck 2018 analysis. The red, blue and magenta curves correspond to a PMF strength of $2.0$ nG,  $2\sqrt{2}$ nG and  $4$ nG, respectively at the comoving length of $1$ Mpc.}
    \label{fig::bb2}
\end{figure}

\begin{figure}[!hbt]
\hspace*{-2cm}
    \centering     
\includegraphics[scale=0.35]{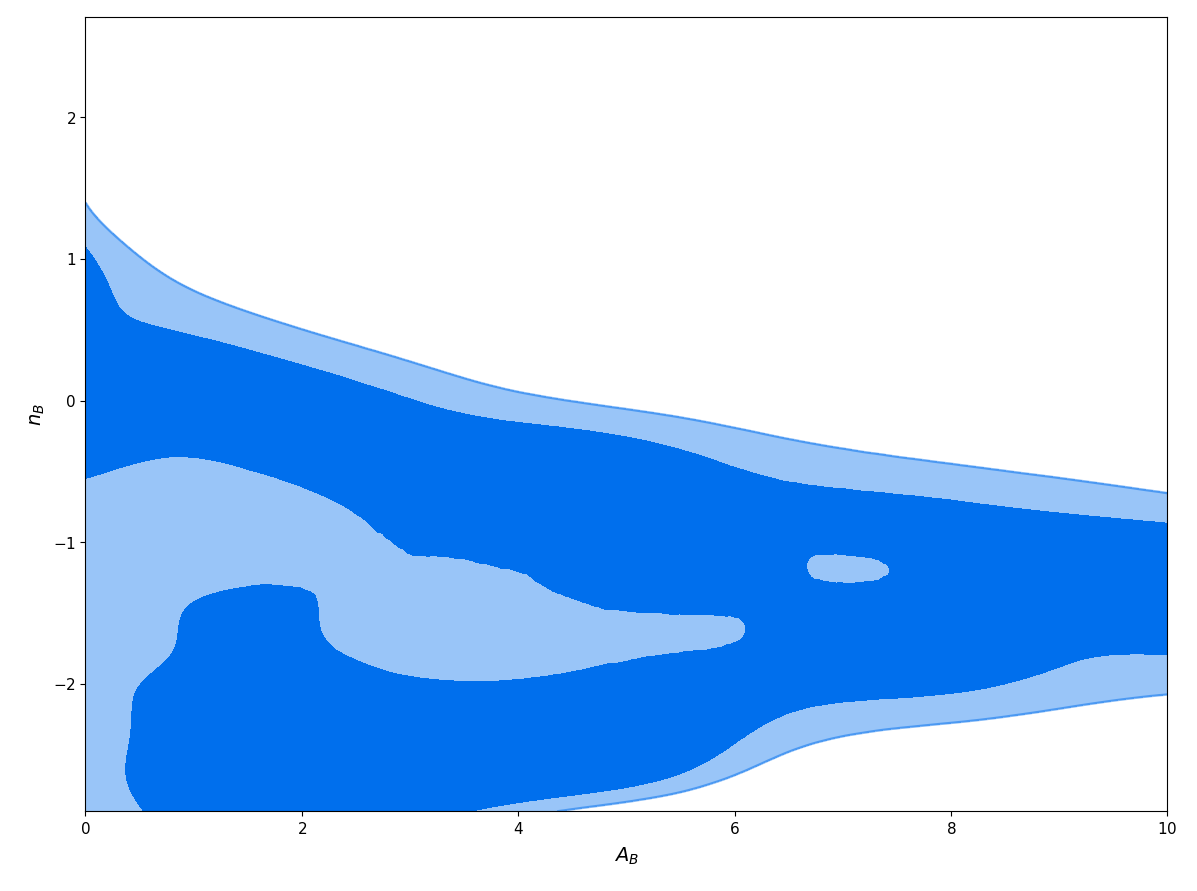}    
    \caption{The figure shows the allowed range of primordial magnetic field parameters in $A_B$ and $n_B$ plane using BICEP2 data. The dark and light blue regions represent $68\%$ and $95\%$ confidence level for the parameters, respectively.}
    \label{fig::bicep_contour}
\end{figure}

\ref{fig::bb2} shows the B-mode spectra along with the BICEP2 and POLARBEAR data with error bars. The different curves show the theoretical prediction of the power spectrum, including the vector, tensor, and lensing contributions. The red, blue, and magenta curves are for non-zero nGauss PMF with increasing strength. The two dashed black curves show the allowed range for BB power-spectra in the absence of primordial magnetic fields. The range of this band is obtained by using the permitted range of different cosmological parameters used in CMB. The range of parameters that are used is consistent with  Planck 2018 constraints \cite{Akrami:2018vks}. Hence, if BB power-spectra lies outside of the band, it is \emph{assuredly not} due to the uncertainty in the values of the cosmological parameters.

We now obtain constraints on the primordial magnetic field strength and the magnetic spectral index from the BICEP2. 
For the analysis, we use the publicly available CosmoMC \cite{Lewis:2002ah} code based on the Metropolis-Hastings algorithm. The code is modified to include the contribution of PMFs to the CMB anisotropies~\cite{Zucca:2016iur}. For the CMB parameters, the priors used are the same as the allowed range of these parameters reported in the Planck 2018 paper. For magnetic field parameters, priors used are: (i) $A_B$ (the amplitude of PMF) with the range $0.1$ to $10$~nG, and (ii) $n_B$ (the magnetic spectral index) in the range $-2.9$ to $3$. The reason for the choice of the parameter $n_B$ to be greater than $-3$ is to avoid the infrared divergence in EMT correlations. Another parameter $log_{10} \tau_{\rm rat}$, which relates neutrino decoupling timing to PMFs generation timing, ranges between 4 to 17.  

\ref{fig::bicep_contour} shows the allowed range of primordial magnetic field strength $A_B$ and its spectral index $n_B$.  The shaded regions show $1\sigma$ and $2\sigma$ contours corresponding to $68\%$ and $95\%$ confidence levels in $A_B$ and $n_B$ plane. 
BICEP2 data alone allows the entire range of $A_B$ considered in the analysis and does not provide any bound on the PMF strength in the nGauss range. However, for the spectral index, we have an upper bound for this range of $A_B$. For $A_B$ $\ge 5$, we also have lower-bound on $n_B$, but for magnetic fields weaker than that, BICEP2 doesn't provide any lower bound on $n_B$. Also, as the magnetic field strength increases, the allowed range for spectral index gets narrower.

\section{Results and conclusions}
\label{sec::results_and_conclusions}

We have shown that a nGauss PMF does not significantly affect the TT, EE, and TE power spectrum. However, the presence of PMF increases the \emph{BB power-spectrum} significantly. For a magnetic field of strength $2$ nG, the BB power-spectrum increases by a factor $2$, and a magnetic field of strength $4$ nG increase BB power-spectrum by a factor $14$ at higher multipoles ($l\sim 2000$). This is mainly because of non-vanishing vector modes which contribute at small scales. The presence of PMF introduces tensor modes as well, which generates BB power-spectrum, we find that the BB power-spectrum is affected at lower multipoles $\ell < 150$. Therefore, the BB power spectrum with PMF shows a significant deviation from that of the vanishing PMF scenario. We have assumed that the neutrinos to be massless. However, the effect of massive neutrinos will not make any difference to the percentage change in the BB spectrum for different $\ell$. Even if neutrinos interact, the interaction is very feeble.

We have shown that the BB power-spectrum with nGauss PMF is higher compared to the BB power-spectrum for the allowed range of parameters from PLANCK 2018 results. Hence, the BB power-spectrum with nGauss PMF is outside of the permitted range and does not arise due to the statistical uncertainty or uncertainty in the measurement of cosmological parameters. We also compared the theoretical predictions with B-mode observations from the BICEP2 data. Although the data is available for multipoles up to $\ell \sim 350$, we find that the power spectra for zero and non-zero PMF are consistent with observations. We constrain the primordial magnetic field parameters using BICEP2 data. The constraints on magnetic field strength $B$ and the spectral index $n_B$ obtained are not as tight as reported in Planck 2015 papers \cite{Ade:2015cva}. BICEP2 data alone doesn't constrain the magnetic field strength in the nano-Gauss range considered in the analysis, but it provides an upper bound on the magnetic spectral index; however, we obtained a lower limit within 2-$\sigma$ error on the spectral index for $A_B \ge 4.5$nG. 

The BICEP2 data can not confirm or infirm the presence of PMF at the epoch of recombination. Since the B-mode power spectra for non-zero PMF shows a significant deviation from the vanishing PMF scenario, the B-mode observations at higher multipoles can detect the presence of nGauss fields. Identifying B-modes requires an instrument with a sensitivity of order $2$ better than the PLANCK mission, and the systematic errors suppressed to a few nKelvin~\cite{2014-Wu-etal-ApJ}. 

Several missions are being proposed to detect the B-modes from CMB, and in the next decade some of them are expected to be operational~\cite{deBernardis:2008bf,Matsumura:2013aja,Ishino:2016izb,2017-CORE-JCAP,Abazajian:2016yjj,Hanany:2019lle}. The aim of B-pol mission~\cite{deBernardis:2008bf} and LiteBird (Lite (Light) Satellite for the studies of B-mode polarization and inflation)~\cite{Matsumura:2013aja,Ishino:2016izb} is to measure the B-modes at the large scales. The aim of CORE (Cosmic ORigins Explorer)~\cite{2017-CORE-JCAP} and PICO (Probe of Inflation and Cosmic Origin)~\cite{Hanany:2019lle} is to obtain the full-sky maps of E-mode and B-mode polarization anisotropies at large and intermediate scales. While the measurement of B-modes in small $\ell$ will provide information about the primordial tensor perturbations and constrain inflation models, our analysis shows that the measurement till $\ell = 500$ will certainly provide a way to detect the primordial magnetic fields. 

The measurement of the BB power spectrum in the 
range $200 < \ell < 500$ is the best scenario for the detection of nGauss
PMF in the future CMB missions. However, even in the pessimistic scenario where the instrument is sensitive up to $\ell < 350$, the future CMB missions can still lead to a strong constraint on the PMF. A 
non-detection of PMF will significantly constrain the early Universe models generating primordial magnetic fields.

\begin{acknowledgments}
The authors wish to thank S. Sethi, T. R. Seshadri, Ajit Srivastava, and K. Subramanian for useful discussions. This work is supported by the ISRO-Respond grant. 
\end{acknowledgments}

%


\end{document}